%% file: main.tex
\documentclass[a4paper, amsfonts, amssymb, amsmath, reprint, showkeys, nofootinbib,superscriptaddress,longbibliography]{revtex4-1}
\usepackage[english]{babel}
\usepackage[utf8]{inputenc}
\usepackage[colorinlistoftodos, color=green!40, prependcaption]{todonotes}
\input{preamble}
\usepackage[pdftex, pdftitle={Article}, pdfauthor={Author}]{hyperref} % For hyperlinks in the PDF

\usepackage{graphicx}% Include figure files
\usepackage{dcolumn}% Align table columns on decimal point
\usepackage{bm}% bold math
\usepackage{braket}
\usepackage{xspace}
\usepackage{enumerate}
\usepackage{booktabs}
\newcommand{\melvin}{{\small M}{\scriptsize ELVIN}\xspace}
\newcommand{\cnot}{{\small C}{\scriptsize NOT}\xspace}

\begin{document}

\preprint{APS/123-QED}

\title{Advances in High Dimensional Quantum Entanglement}% Force line breaks with \\
\author{Manuel Erhard}
\email{manuel.erhard@univie.ac.at}
\affiliation{Vienna Center for Quantum Science \& Technology (VCQ), Faculty of Physics, University of Vienna, Austria.}
\affiliation{Institute for Quantum Optics and Quantum Information (IQOQI) Vienna, Austrian Academy of Sciences, Austria.}
\author{Mario Krenn}
\email{mario.krenn@univie.ac.at}
\affiliation{Vienna Center for Quantum Science \& Technology (VCQ), Faculty of Physics, University of Vienna, Austria.}
\affiliation{Institute for Quantum Optics and Quantum Information (IQOQI) Vienna, Austrian Academy of Sciences, Austria.}
\affiliation{Department of Chemistry \& Computer Science, University of Toronto, Canada.}
\affiliation{Vector Institute for Artificial Intelligence, Toronto, Canada.}
\author{Anton Zeilinger}
\email{anton.zeilinger@univie.ac.at}
\affiliation{Vienna Center for Quantum Science \& Technology (VCQ), Faculty of Physics, University of Vienna, Austria.}
\affiliation{Institute for Quantum Optics and Quantum Information (IQOQI) Vienna, Austrian Academy of Sciences, Austria.}

\date{\today}% It is always \today, today,

\begin{abstract}
Since its discovery in the last century, quantum entanglement has challenged some of our most cherished classical views, such as locality and reality. Today, the second quantum revolution is in full swing and promises to revolutionize areas such as computation, communication, metrology, and imaging. Here, we review conceptual and experimental advances in complex entangled systems involving many multilevel quantum particles. We provide an overview of the latest technological developments in the generation and manipulation of high-dimensionally entangled photonic systems encoded in various discrete degrees of freedom such as path, transverse spatial modes or time/frequency bins. This overview should help to transfer various physical principles for the generation and manipulation from one to another degree of freedom and thus inspire new technical developments. We also show how purely academic questions and curiosity led to new technological applications. Here fundamental research provides the necessary knowledge for coming technologies such as a prospective quantum internet or the quantum teleportation of all information stored in a quantum system. Finally, we discuss some important problems in the area of high-dimensional entanglement and give a brief outlook on possible future developments. 
\end{abstract}

\maketitle

%\tableofcontents

\tableofcontents

\section{Introduction}
The basic unit of quantum information theory and technology is the qubit, which is based on a two-state system. Nature, however, uses a four-letter alphabet for the arguably most important storage system, the DNA. Although it is still unclear why evolution developed a four-letter system\cite{szathmary2003there}, it may give us an indication that we should also look at more complex systems than the one based on bits.

\begin{figure*}[t!]
\centering
\includegraphics[width=.8\linewidth]{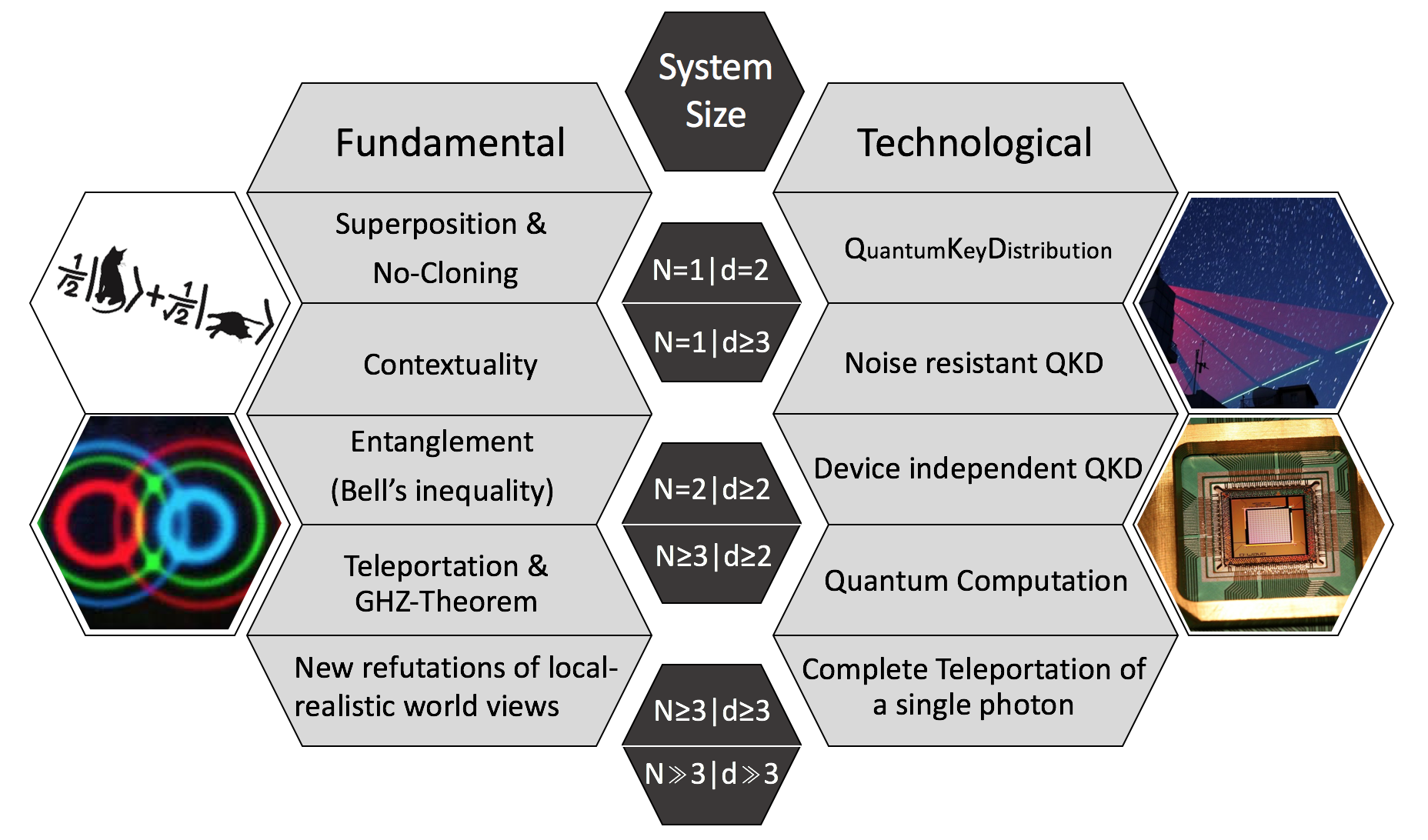}
\caption{\textbf{Quantum paradoxes arising from system size.} The left of the two columns shows the curiosity-driven fundamental questions and theorems with an increasing number of involved particles (N) and their respective local dimensionality (d). Surprisingly, these findings inspired actual technological applications that will power the second quantum revolution, displayed in the right column. QKD, Quantum-Key-Distribution, GHZ, Greenberger-Horne-Zeilinger}
\label{fig:fundamental-technology-in-terms-of-size}
\end{figure*}

In fact, it is known in quantum communication that protocols based on larger alphabets offer certain advantages. In addition to a higher information capacity especially the increased resistance to noise is interesting for applications\cite{bechmann2000quantum,cerf2002security}. Also relaxed constraints in fundamental tests of nature, such as the violations of local-realistic theories, are known in higher-dimensional systems\cite{vertesi2010closing}.

Several physical systems allow encoding of higher dimensional quantum information. These systems are as diverse as Rydberg atoms, trapped ions\cite{senko2015realization}, polar-molecules\cite{yan2013observation}, cold atomic ensembles\cite{parigi2015storage,ding2016high}, artificial atoms formed by superconducting phase qudits\cite{neeley2009emulation}, defects in solid states\cite{choi2017observation} or photonic systems. Here we want to focus specifically on discretized degrees of freedom (DoF) of photons, such as path, transverse-spatial modes or time/frequency bins. Information on  photonic continuous variable systems can be found in\cite{braunstein2005quantum,weedbrook2012gaussian}. 

The focus is on experimental techniques for the generation and manipulation of high-dimensional entanglement. We explain the physical principles of various methods to create high-dimensional entanglement and present the state of current experimental efforts that realize different approaches. In general, this review aims to shed light on various DoFs, highlighting analogies and possible synergies between the techniques used in laboratories around the world. We hope that ideas from various DoFs will inspire other physicists to come up with new ideas on how to create and manipulate high-dimensional entanglement.

Theoretical basics and current state-of-the-art methods in characterizing and detecting complex quantum entanglement can be found in these excellent reviews\cite{bruss2002characterizing,horodecki2009quantum,guhne2009entanglement,plenio2014introduction,friis2018entanglement}.

\section{New possibilities by increasing the number of particles and dimensions}
It is interesting from a historical perspective that high-dimensional entangled quantum systems have been considered theoretically from the very beginning in 1935 started by Einstein, Podolsky, Rosen (EPR) and later Schrödinger\cite{einstein1935can, Schrodinger:1935cq}. They considered the external and continuous parameters of quantum systems, namely the position and momentum of two strongly correlated (entangled) particles. Only in 1951, David Bohm \cite{bohm1951quantum} came up with the idea to investigate two entangled spin-$\frac{1}{2}$ particles, i.e. two-state systems called qubits. The seminal work by J.S. Bell in 1964\cite{bell1964einstein} moved this formally purely philosophical questions about entanglement, reality and locality to an experimentally testable theorem. These developments aroused the interest of many experimental physicists who in the following years attempted to generate entangled quantum systems in the laboratory\cite{freedman1972experimental,aspect1982experimental,ou1988violation,brendel1992experimental,kwiat1993high,kwiat1995new}.

From this brief historical point of view, it seems that going from complex systems to more simple ones allowed for several important observations in quantum mechanics. We believe that this is true, especially if we keep in mind that discovering and appreciating these mind-boggling ideas is much easier in the most simplified setting. However, we would like to change the perspective from a chronological viewpoint to one where the system size in terms of numbers of particles and their local dimensionality (complexity) are the driving factors. As shown in Fig.\ref{fig:fundamental-technology-in-terms-of-size}, we also introduce two categories: Fundamental and Technological. We mention these two categories to emphasize that many applications of quantum technology have their origins in fundamental and initially purely academic questions.
                                   
The principle of superposition already contains the essential features of quantum mechanics\cite{feynman_quantum_1965}. The famous double-slit experiment gives the prime example of the superposition principle. If we consider a single electron incident on a double-slit, then after transmitting through the double-slite, the quantum state is in a superposition of the two possible ways the electron could have taken. If we now ask whether we can simply copy this unknown quantum state, the surprising answer is no\cite{wootters1982single}. At first glance, the \textit{no-cloning} theorem seems like a very strong limitation. This is true especially if we compare to classical information science, where the ability to copy information is essential for long-distance communication or error-correcting schemes. Surprisingly, the superposition principle and the no-cloning theorem leads to a useful application, namely quantum key distribution (QKD)\cite{bennett2014quantum}. In QKD schemes, a key is distributed between two parties in a secure way such that if an eavesdropper reveals too much information about the key it can be detected. 
Single quantum particles in higher dimensions result in questions about possible representations using hidden variables models\cite{kochen1975problem}. Here the contextuality of quantum measurements and its corresponding hidden value can be investigated. Interestingly, there is already a strong contradiction between quantum and classical physics, without even considering entanglement. It was found in quantum mechanics, that the measurement outcome depends on the measurement performed, meaning the context of the measurement itself\cite{lapkiewicz2011experimental}. On the technological side, increasing the dimensionality of a single particle not only leads to higher information capacities, but more importantly to unprecedented levels of noise-resistance in QKD schemes\cite{bechmann2000quantum,cerf2002security}.

Going from single to two locally separated quantum systems, questions regarding locality had enormous influence on our worldview. Schrödinger coined the term \textit{entanglement} to describe non-classically correlated quantum systems and described it as \textit{``... not one but rather the characteristic trait of quantum mechanics''}\cite{schrodinger1935discussion}. Philosophically and from fundamental importance is whether quantum mechanics is compatible with local-realistic hidden variable theories as considered by Einstein-Podolsky-Rosen (EPR) \cite{einstein1935can} and Bell\cite{bell1964einstein}. A series of recent experiments closing all essential experimental loopholes denied this\cite{hensen2015loophole,giustina2015significant,shalm2015strong}. 
A surprising twist in history yielded the investigations into generalized Bell-like violations in higher-dimensional bipartite systems. The first theoretical investigation showed classical behavior for high spin values\cite{mermin1980quantum}. Later, larger violations of local-realistic theories have been found than in the qubit case which also unveiled a higher noise resistance as dimensionality increased\cite{kaszlikowski2000violations,durt2001violations,collins2002bell}.

Although entanglement itself does not allow the transmission of information, it can serve as the ultimate cryptography channel. Bell-type violations in combination with classical communication  enable device-independent concepts of key distribution for cryptography \cite{ekert1991quantum, mayers1998quantum}.

Multi-particle systems with particle numbers $N$ greater than two do not only increase the Hilbert space exponentially but also lead to qualitatively new insights into the relationship between classical and quantum physics. One such example is the Greenberger-Horne-Zeilinger (GHZ) theorem\cite{greenberger1989going}. The essential difference between Bell's and GHZ theorem is that in case of GHZ the predicted outcomes by the quantum theory are deterministic. Thus the assignment of a local hidden variable is more direct from a conceptual point of view. On the application side, many qubits result in the possibility of universal quantum computation. Also since physical calculators are never perfect and classical error-correcting schemes cannot be applied to quantum computing due to the no-cloning theorem, multi-particle entanglement in the form of GHZ states can be used for quantum error correction schemes\cite{shor1995scheme}.

While the original GHZ argument for three and more particles in two dimensions was formulated more than two decades ago~\cite{greenberger1989going,mermin1990extreme}, its generalization to arbitrary local dimensions was only found recently~\cite{ryu2013greenberger,PhysRevA.89.024103,lawrence2014rotational,tang2017multisetting}. A key difference to the two-dimensional qubit case is that the N-partite observables do not form a commuting set of observables and are not hermitian. However, all observables have the multi-partite and high-dimensional GHZ state as a common eigenstate\footnote{A set of such observables is called a concurrent set.}, thus still predict an outcome with certainty. Most interestingly, so far no local-realistic violation of the GHZ type could be constructed using local hermitian operators. All attempts to generalize the GHZ argument till date use local unitary observables and thus have complex eigenvalues. This is in sharp contrast to the prevailing view in physics that physical observables must have real eigenvalues, as famously stated by Dirac\cite{dirac1981principles}. This fact calls for a more general definition of what properties a physical observable should obey. A more general class of operators that fulfil the requirement of orthogonal eigenstates are normal operators\cite{hu2017observables}. 

If the system size in terms of number and dimensionality grows to extensive numbers, exotic phenomena arise. Examples are superconductivity, super-fluids or Bose-Einstein condensates. These systems still pose significant theoretical as well as experimental challenges. A deeper understanding of these extremely large and highly correlated quantum systems might very well reveal new physics.

\begin{figure*}[htp]
	\centering
	\includegraphics[width=0.75\linewidth]{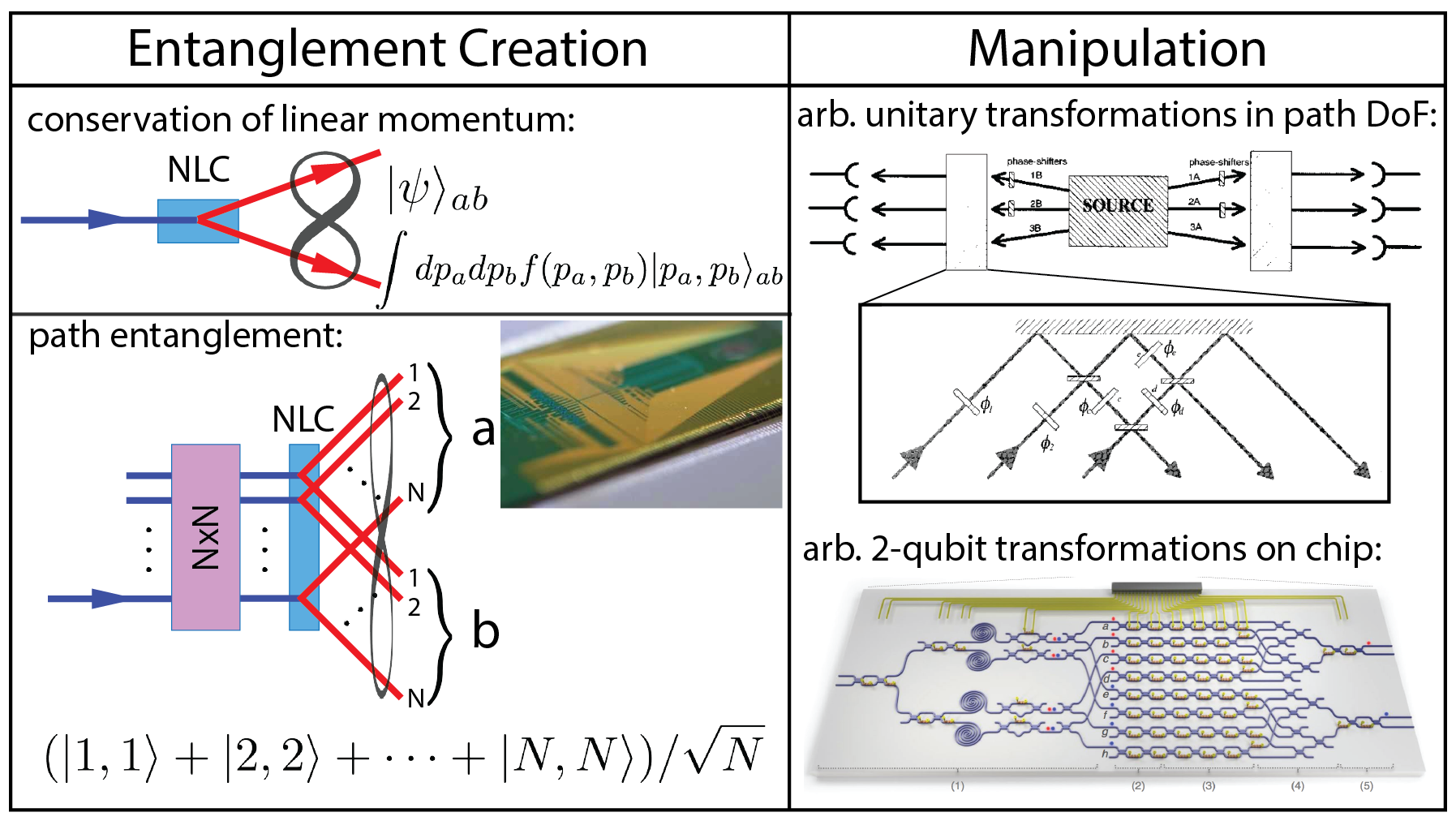}
	\caption{\small{\textbf{Entanglement creation and manipulation concepts for path DoF.} Due to the conservation of linear momentum in a spontaneous parametric down-conversion process in a non-linear crystal (NLC) a high-dimensionally entangled two-photon state $\ket{\psi}_{ab}$ is created. A conceptually different method to create path-entangled quantum states is to utilize a $N\times N$ multi-port that coherently splits a pump beam into N beams. These N coherent pump beams can now each create a photon pair in their respective paths $a_i$ and $b_i$ with $i\in\{1,\cdots,N\}$. Keeping the probability of creating a photon pair much smaller than one results in the creation of an N-dimensional quantum state $\ket{\psi}_{ab}$.
	The path DoF offers a unique feature regarding the coherent manipulation of quantum information. Using the scheme introduced by Reck and Zeilinger\cite{reck1994experimental}, which only consists of beam splitters and phase shifters, any arbitrary unitary transformation can be implemented. Both, the entanglement creation and manipulation techniques are perfectly suited for on-chip implementations as indicated by the small inset, image taken from University of Bristol. Even arbitrary 2-qubit transformations necessary for quantum computation have been implemented on chip as depicted above, figure taken from\cite{qiang2018large}.}}
	\label{fig:path-unitary-and-entanglement-creation}
\end{figure*}
\section{Photonic carriers of high-dimensional entanglement}
In this section, we discuss different physical and technical methods of how to generate and manipulate higher-dimensional entangled pairs of photons. Besides, we also investigate the experimental ability to detect high-dimensional entanglement. We focus on local measurements as they are interesting from a fundamental as well as an application point of view. Since photons are the ideal carriers of quantum information over long distances, we also present current approaches to distribute the quantum information stored in different DoFs.
\\
Despite the very important theoretical and experimental effort to deterministically generate single and multiple photons, a discussion of these methods would go beyond the scope of this review. Moreover, many of the schemes to create entangled photon pairs in two or higher-dimensions, actually rely on this inherent probabilistic nature of the photon source. Hence, in the following, we will focus on probabilistic photon-pair sources such as spontaneous parametric down-conversion (SPDC) and spontaneous four-wave mixing (SFWM). 

SPDC is based on a $\chi^{(2)}$ non-linear coefficient of some material. The phase-matching conditions are given by
\begin{eqnarray}\label{eq:spdc-phase-matching}
\centering
	\omega_p&=\omega_i+\omega_s\\\nonumber
	k_p&=k_s+k_i,
\end{eqnarray}
with $\omega$ describing the photons frequency, $k$ the linear momentum and $p,i,s$ refer to pump, idler and signal, respectively.
The first equation in eq.~\ref{eq:spdc-phase-matching} stems from energy conservation\footnote{since the energy $E$ of a photon is given by $E=\hbar\omega$} and the second equation from conservation of linear momentum. Satisfying the phase-matching condition results in the spontaneous (i.e. completely probabilistic) down-conversion of a pump photon into two photons according to eq.~\ref{eq:spdc-phase-matching}. In the low pump power regime, the quantum state of the down-converted photon pair can be written as a Taylor expansion\cite{ghosh1986interference,mandel1995optical}
\begin{equation}\label{eq:photon-pair-approx-fock-basis}
	\ket{\psi}\approx\ket{0,0}+\alpha\ket{1,1}+\alpha^2/2\ket{2,2}+\cdots,
\end{equation}
where we used the photon number basis (Fock-basis). Due to the inherently probabilistic emission of photon pairs with probability amplitude $\alpha$, there is also the probability that two or more photon pairs are emitted simultaneously. Thus $\alpha$ is usually set much smaller than one $\alpha\ll 1$.
\\
In contrast, SFWM is based on the $\chi^{(3)}$ non-linearity and the corresponding phase-matching conditions are
\begin{eqnarray}\label{eq:sfwm-phase-matching}
\centering
	&2\omega_{p}=\omega_i+\omega_s\\\nonumber
	&2 k_{p}=k_s+k_i,
\end{eqnarray}
using the same notation as for SPDC but now with two pump fields $p$. Thus in SFWM two pump photons are converted to two output photons which can also be described by equation~\ref{eq:photon-pair-approx-fock-basis}.

Several specific advantages arise using either SPDC or SFWM for different implementations and applications, as we will discuss in the following sections. A thorough review of single-photon sources, as well as single-photon detectors, can be found here\cite{eisaman2011invited}.

\subsection{Path}
Quantum information encoded in the path DoF is appealing because there exist arbitrary single-photon transformations in any dimension\cite{reck1994experimental}. This scheme only utilizes beam splitters and phase-shifters and allows to implement any single qudit transformation, see Fig.\ref{fig:path-unitary-and-entanglement-creation}a). In addition, the experimental realisation is possible using bulk optical elements or integrated optics, e.g. silicon chips with extreme interferometric stability and high indistinguishability\cite{wang2018multidimensional,schaeff2015experimental}, see Fig.\ref{fig:path-unitary-and-entanglement-creation}c). Conceptually, one method of creating high-dimensionally entangled photon pairs relies on the intrinsic momentum conservation within the SPDC process\cite{zukowski1997realizable}. In this scheme, the intrinsic linear momentum conservation leads to the coherent emission of a single photon pair on a cone. The momentum conservation is responsible for the diametrically opposite positions on the emission cone of the two photons. Collecting these photon pairs with $d$ single-mode fibre pairs arranged evenly spaced on the emission cone leads to a $d$-dimensionally entangled quantum state of two photons in their respective fibres or paths\cite{rossi2009multipath}. 

A direct way to harness the high-dimensionally entangled photon pairs in their linear momentum (as depicted in Fig.\ref{fig:path-unitary-and-entanglement-creation}) is to use a deformable mirror device in combination with a single-photon detector. This leads to pixel entanglement\cite{o2005pixel}. In this scheme, the basis transformation for certifying entanglement can be performed using a single lens only, since a lens effectively performs a Fourier-transformation and thus transforms between the position and momentum bases. Efficient detection methods using sparse-matrix techniques allowed to certify a channel capacity of 8.4 bits per photon\cite{howland2013efficient}.

A conceptually different method to create high-dimensionally entangled photon pairs is to utilize $d$-indistinguishable photon-pair sources\cite{schaeff2012scalable,luo2019quantum}, see Fig.\ref{fig:path-unitary-and-entanglement-creation}. The physical principle here is based on coherently pumping $d$ NLCs which results in a coherent superposition of a single-photon pair emitted in one of the paths. The crucial ingredient here is to have $d$ indistinguishable photon-pair sources in all DoFs except the path, where the quantum information is stored. Experimentally, this scheme can either be realized using bulk optical elements\footnote{This approach guarantees higher efficiencies >99.9\% due to special anti-reflection coatings, but is difficult to stabilize and scale to very high-dimensions.}, integrated fibre optical elements or on-chip techniques, see inset in  Fig.\ref{fig:path-unitary-and-entanglement-creation}. 

Particularly interesting are the technical advances in recent years in on-chip photonics. This technology is a promising platform for quantum optics experiments that encode the quantum information in the path DoF. High-quality interferometers\cite{harris2017quantum}\footnote{Extinction ratios up to 66dB are reported in the literature.} and scalability in the number of optical components allow a variety of applications. Probabilistic processes based on either spontaneous four-wave mixing (SFWM)\cite{sharping2006generation} or spontaneous parametric down-conversion (SPDC)\cite{jin2014chip}\footnote{Here Lithium-Niobate waveguides are used, for example.} are used as photon-pair sources. SFWM has the advantage that all materials used (e.g. Silicon or Hydex) are complementary metal–oxide semiconductor (CMOS) compatible. A common source of photon pairs is the SFWM based spiral waveguide source, where the phase-matching conditions can be achieved by appropriate design of the spiral waveguides through anomalous dispersion.

Sources of this type have recently been used to certify three\cite{xiasong2019chip3d} and 14-dimensional entanglement on-chip in the path DoF\cite{wang2018multidimensional}. In the latter, a total of 16 identical spiral waveguide sources based on SFWM with a total of more than 550 optical components were implemented on-chip. The observed fidelities range from $96\%$ for $d=4$ to $81\%$ for $d=12$. Typically, each of the 16 sources produces photon pairs at a rate of 2kHz.

Another experiment recently demonstrated an optical general 2-qubit quantum processor implemented on-chip\cite{qiang2018large}. There, two pre-entangled ququarts are used to probabilistically perform any 2-qubit unitary operation, such as \cnot gates for example, between two qubits encoded in the path DoF. Also quantum Hamiltonian learning algorithms\cite{wiebe2014hamiltonian} have been performed using higher-dimensionally encoded path qudits on chip\cite{wang2017experimental}. In their work, a negatively charged nitrogen-vacancy (NV) center has been coupled to a trusted quantum simulator, in this case the photonic silicon chip. This chip was then used to learn the Hamiltonian of the NV center.

\begin{figure*}[htp]
	\centering
	\includegraphics[width=0.65\linewidth]{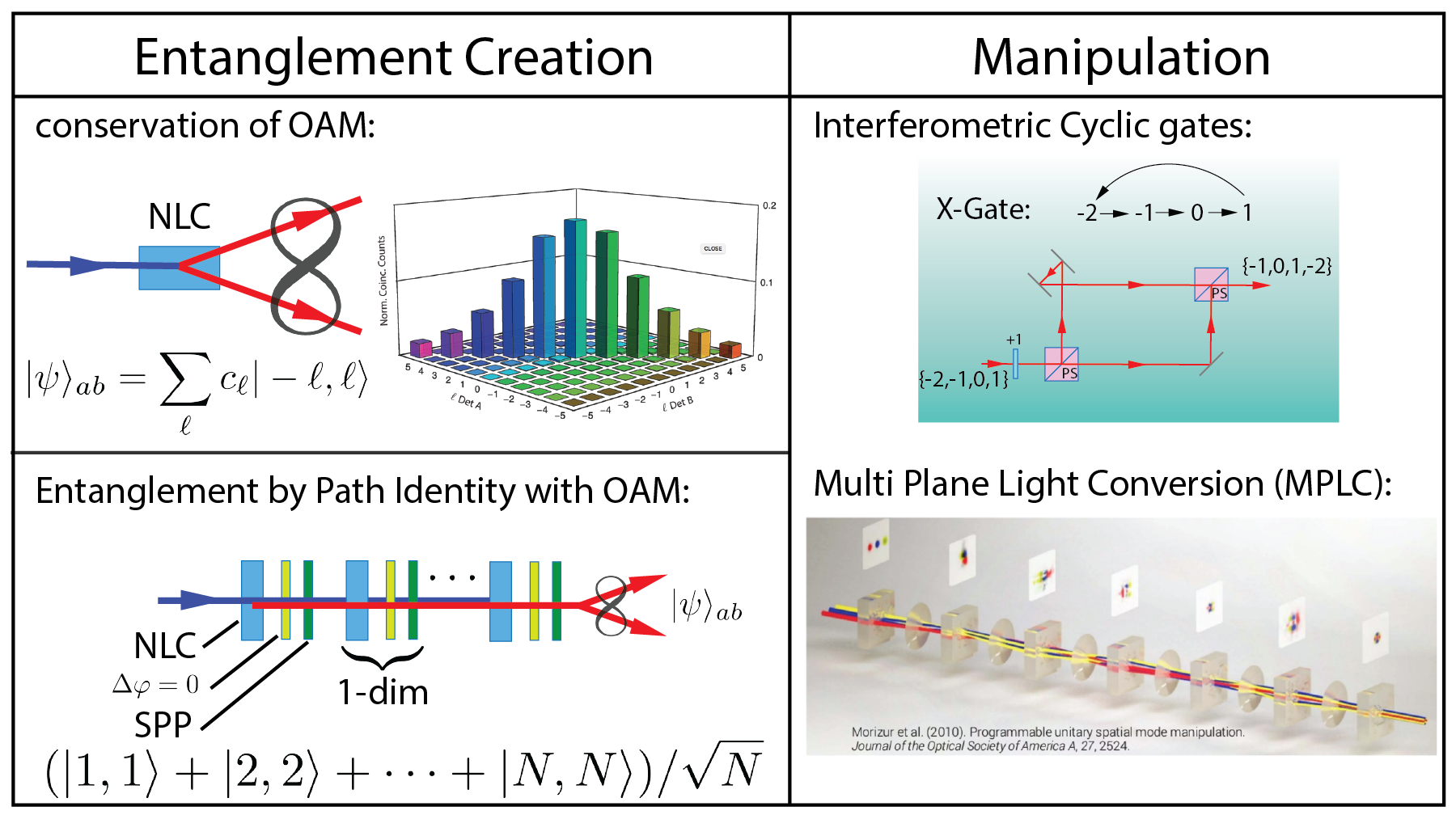}
	\caption{\small{\textbf{Entanglement creation and manipulation concepts for orbital-angular-momentum (OAM) DoF.} Approximate conservation of OAM within the collinear spontaneous-parametric down-conversion process in a non-linear crystal (NLC) leads up to 100-dimensionally entangled two-photon quantum states\cite{krenn2014generation} $\ket{\psi}_{ab}$. The unequal coefficients $c_\ell$ usually lead to non-maximally entangled states, as shown in the inset taken from\cite{erhard2017quantum}.
	The concept of \textit{entanglement by path-identity} relies on $N$ indistinguishable photon-pair sources (NLCs) \cite{krenn2017entanglement,krenn2017quantum}. Each dimensionality unit (1-dim) contains a NLC, a phase-shifter (which is set to zero here for simplicity $\Delta\varphi=0$) and a mode-shifter in the form of a spiral-phase-plate (SPP) that is capable of adding one quantum of OAM to incoming photons. Stacking N such 1-dim modules coherently yields a maximally entangled two-photon quantum state as described in the figure. Some basic high-dimensional quantum gates, such as cyclic-gates, can efficiently be performed for OAM. These methods rely on the parity sorter (PS), which is capable of sorting incoming photons with even/odd amount of OAM into different paths\cite{leach2002measuring}. A recently introduced and conceptually very different approach, called multi-plane-light-conversion (MPLC)\cite{morizur2010programmable} promises arbitrary unitary transformations for OAM. The principle is to utilize several phase-holograms with propagation in between to achieve mode-matching between input and output modes, which can be defined arbitrarily.}}
	\label{fig:oam--entanglement-creation-mplc-xGate}
\end{figure*}

Distributing the classical information using the path DoF is conveniently possible using multi-core-fibres (MCF). These MCFs find applications in classical communication networks, specifically where multiplexing schemes for higher information capacities are necessary. However, transferring quantum information using MCFs is more involved, since the phase between different cores must be interferometrically stable to avoid dephasing. Random dephasing results in incoherent and thus classical mixtures which yield non-entangled quantum states. However, recent results achieved four-dimensional quantum-key-distribution using MCFs of up to 0.3km\cite{canas2017high,ding2017high}. Also, high-dimensional entanglement has been transmitted through two MCFs recently~\cite{lee2017experimental}. 
An interesting approach is to send pixel entanglement via a conventional graded-index multi-mode fiber\cite{valencia2019unscrambling}. Using entanglement itself to measure the complex transmission matrix, scrambling of the optical modes within the optical fiber can be compensated by adjusting the measurement bases accordingly. So far, 6-dimensional entanglement over a length of 2m can be sent. These proof-of-principle demonstrations show the possibilities to use MCFs for high-dimensional quantum information to connect optical chips\cite{ding2019demonstration}, in future even in high-dimensions.

\subsection{Transverse Spatial modes of photons}
The transverse spatial modes of single photons can perfectly encode high-dimensional quantum information. Different mode-families have been studied in the context of entanglement including Laguerre-Gauss\cite{mair2001entanglement}, Bessel-Gaus\cite{mclaren2012entangled} and Ince-Gauss\cite{krenn2013entangled} modes. Here, we choose to solely discuss the Laguerre-Gaussian (LG) modes\cite{allen1992orbital}, since the majority of the experimental work regarding high-dimensional entanglement has been performed with the orbital-angular-momentum (OAM) of photons\cite{erhard2018twisted}. The OAM describes the transverse wavefront of photons. The main feature of these modes is the singularities within the phase\cite{yao2011orbital,molina2007twisted}. The amount of OAM in $\hbar$ corresponds to the direction and number of windings $\ell$ of the phase around these singularities. At the singularity, the phase is not defined, which results in the typical doughnut-shaped intensity distributions. The reason why OAM is interesting for quantum experiments is two-fold: First, OAM entanglement can easily be created in a single NLC, because the OAM is conserved within the SPDC process\cite{mair2001entanglement} and thus directly yields high-dimensionally entangled photon pairs\cite{vaziri2002experimental}. Secondly, there are several experimentally feasible techniques known how to manipulate and measure OAM states of single photons\cite{mair2001entanglement,leach2002measuring,karimi2014generating}. 

A quantum state of two photons generated in a NLC can be described according to $\sum_{\ell} c_\ell\ket{\ell,-\ell}_\text{ab}$, where $\ell$ describes the amount of OAM in $\hbar$ and the dimensionality is given by $d=2|\ell_\text{max}|+1$. The exact distribution of the complex coefficients $c_\ell$ is called \textit{spiral-spectrum} and mainly depends on the length of the crystal, the beam waist of the pump laser and the collection beam waist\cite{miatto2012bounds}. It is even possible to optimize the entanglement dimensionality using the phase matching conditions\cite{romero2012increasing}. Using OAM entangled photon pairs created in a single NLC, high-dimensional generalized Bell inequalities have been violated up to 12-dimensions\cite{dada2011experimental}. Including the complete LG-modes, namely $\ell, p$-modes\footnote{The LG-modes are represented by two indices: $\ell$ denotes the azimuthal phase of the topological charge in terms of OAM and $p$ represents the radial quantum number\cite{karimi2014radial,karimi2014exploring,plick2015physical}.}, of the two-dimensional transverse spatial field of the photons into the entangled two-photon quantum states can lead to more than 100-dimensional entanglement in the laboratory\cite{krenn2014generation}. Creating maximally entangled\footnote{i.e. $|c_\ell|^2=1/d~\forall~\ell$} states can be either achieved using \textit{procrustean} filtering techniques\cite{vaziri2003concentration} or taking the natural spiral-spectrum of the SPDC into account and counteract with a corresponding superposition of different OAM quanta in the pump beam~\cite{kovlakov2018quantum,liu2018coherent}.

Recently, a conceptually new method for creating OAM entangled quantum states relying on indistinguishability and path-identity has been introduced\cite{krenn2017entanglement,kysela2019experimental}. In this approach, see Fig.\ref{fig:oam--entanglement-creation-mplc-xGate}, $d$ NLCs are pumped coherently, and their respective paths are identically aligned, such that the resulting quantum state is in a coherent superposition of one photon-pair being emitted in one of the $d$-NLCs. To create an OAM entangled quantum state, a spiral-phase-plate (SPP)\footnote{A SPP is a device that adds $m$ quanta of OAM to the incoming photons, thus realizing the operation $\ket{\ell}\rightarrow\ket{\ell+m}$. Q-plates are perfectly suited for this task, especially in the collinear regime\cite{marrucci2006optical}.} is inserted after each NLC. Depending on the relative pump powers and phases\footnote{The pump power (relative phase) given by $|c_\ell|~(\text{exp}(i\varphi_\ell))$ controls the magnitude and phase of every coefficient within the entangled quantum state $c_\ell=|c_\ell| \text{exp}(i\varphi_\ell)$} between the NLCs an arbitrary $d$-dimensionally entangled two-photon quantum state is created. Currently, fidelities for three-dimensional entanglement of approximately $F\approx 90\%$ is achieved. Future efforts in integrating the NLCs directly with q-plates could pave the way to create tens of dimensions with even higher fidelities.

Manipulation of OAM in higher-dimensions has been proven difficult. Although projective measurements using phase-plates or fork holograms in combination with single-mode fibres are well established\cite{heckenberg1992generation,mair2001entanglement} and provide a powerful tool to detect high-dimensional entanglement in only two non-orthonormal bases\cite{bavaresco2018measurements}, even seemingly simple transformations such as cyclic transformations were not known until recently\cite{schlederer2016cyclic,babazadeh2017high} and have only been discovered using the computer algorithm \melvin \cite{krenn2016automated}. At the heart of these efforts lies a simple but very elegant device that is capable of sorting OAM quanta according to their parity (even/odd)\cite{leach2002measuring}. Based on a Mach-Zehnder interferometer with a Dove-Prism inserted that introduces an OAM dependent phase, different parity OAM values can be sorted. Cascading the parity sorter leads to the capability of sorting arbitrarily many different OAM modes and thus enables multi-outcome measurements. Interestingly, to construct $d$-level cyclic transformations only $\text{log}(d)$ parity sorters are necessary\cite{gao2019arbitrary}. Analogously, this concept has been used to sort different $p$-modes and thus yield access to the complete two-dimensional transverse spatial wave-front of single photons\cite{gu2018gouy,fu2018realization,zhou2019using}. Parity sorters for OAM also allow to route high-dimensionally entangled states\cite{erhard2017quantum} and plays a crucial role in creating genuinely high-dimensional and multi-photon entangled quantum states\cite{malik2016multi,erhard2018experimental}, see Chapter 4 for more details. 

An alternative method to sort different OAM states is based on log-polar transformations, which essentially converts the circularly varying phase pattern to a linear grating\cite{berkhout2010efficient}. This results in the conversion of OAM to linear momentum, and thus different OAM modes appear in different positions (paths) after propagation. Mode sorters for mode numbers up to 50 have been demonstrated~\cite{lavery2013efficient} with next mode overlap of approximately $2\%$\cite{mirhosseini2013efficient}.

A recently developed and demonstrated technique is based on multi-plane light conversion (MPLC)\cite{fontaine2017design,morizur2010programmable}. Using several consecutive phase planes with propagation in between allows transforming arbitrary two-dimensional transverse light fields unitarily. With this approach, 210 modes defined in the LG-modes can be sorted into a spatial pattern of Gaussian spots\cite{fontaine2019laguerre}. In terms of efficiency, the device has a theoretical insertion loss of 2.5dB and a measured insertion loss of 6dB. The difference arises mainly due to reflection inefficiencies of the SLM\footnote{Here, in total seven reflections on an SLM with 0.5dB per reflection is utilized.}. The average crosstalk per mode is measured to yield -30dB, which results in a measured channel capacity of 6.25 bits/photon (theoretically expected is $\text{log}_2(210)=7.71$). MPLC is not only limited to sorting LG-modes but can also be applied to arbitrary unitary transformations, such as cycling operations or controlled operations on a single-photon level\cite{brandt2019high}. This promising route towards complete control of qudits encoded in the LG-modes showed a high process purity of 99\% for three-dimensional cyclic gates using three-phase planes at the SLM.

There are several demonstrations of distributing classical and quantum information using the OAM DoF. Free-space links\cite{gibson2004free} allowed the transmission of classical information in high-speed terabit configurations\cite{wang2012terabit,bozinovic2013terabit}, turbulent intra-city links\cite{krenn2014communication}, underwater channels\cite{bouchard2018quantum} or over large distances up to 143km between two islands\cite{krenn2016twisted}. Also, specifically designed optical fibres have been employed to transmit classical\cite{ramachandran2009generation,wong2012excitation,brunet2014design,brunet2015design,gregg2015conservation,willner2015optical} and quantum information\cite{sit2018quantum} with OAM. High-dimensional quantum key distribution using OAM has been demonstrated in turbulent environments such as intra-city free-space links\cite{sit2017high}, fibre based systems\cite{cozzolino2019orbital} and even entanglement distribution\cite{krenn2015twisted}. 

In addition to all the properties mentioned above, OAM is also perfectly suited for fundamental tests in quantum mechanics regarding two-dimensional entanglement incorporating very high angular momenta quanta of up to $10.010\hbar$\cite{fickler2012quantum,fickler2016quantum}. In these experiments, the question of whether a \textit{macroscopic} bound in terms of the amount of electromagnetic action $\hbar$ exists. Experiments up to $10.010\hbar$ show that there only seems to exist a technical, not a fundamental limitation. Also, the appealing patterns formed by different OAM superpositions of entangled photons\cite{fickler2013real}\footnote{see video https://www.youtube.com/watch?v=wGkx1MUw2TU} or remotely prepared quantum states\cite{erhard2015real} has been imaged live.

\subsection{Discretized Time and Frequency modes}
Quantum information stored in time-bins or in the frequency degree of freedom of single photons is perfectly suited for transmission over large distances using free-space links or optical fibres\cite{tittel1998violation,marcikic2004distribution,steinlechner2017distribution,kues2017chip}. The idea of using time-bins as the physical carrier of quantum information goes back to Franson in 1989 who discussed it theoretically in terms of violating Bell inequalities~\cite{franson1989bell}. A series of experiments demonstrated two-dimensional time-bin entanglement\cite{tittel1998violation,brendel1999pulsed,tittel2000quantum}. 

\begin{figure*}[t!]
	\centering
	\includegraphics[width=0.65\linewidth]{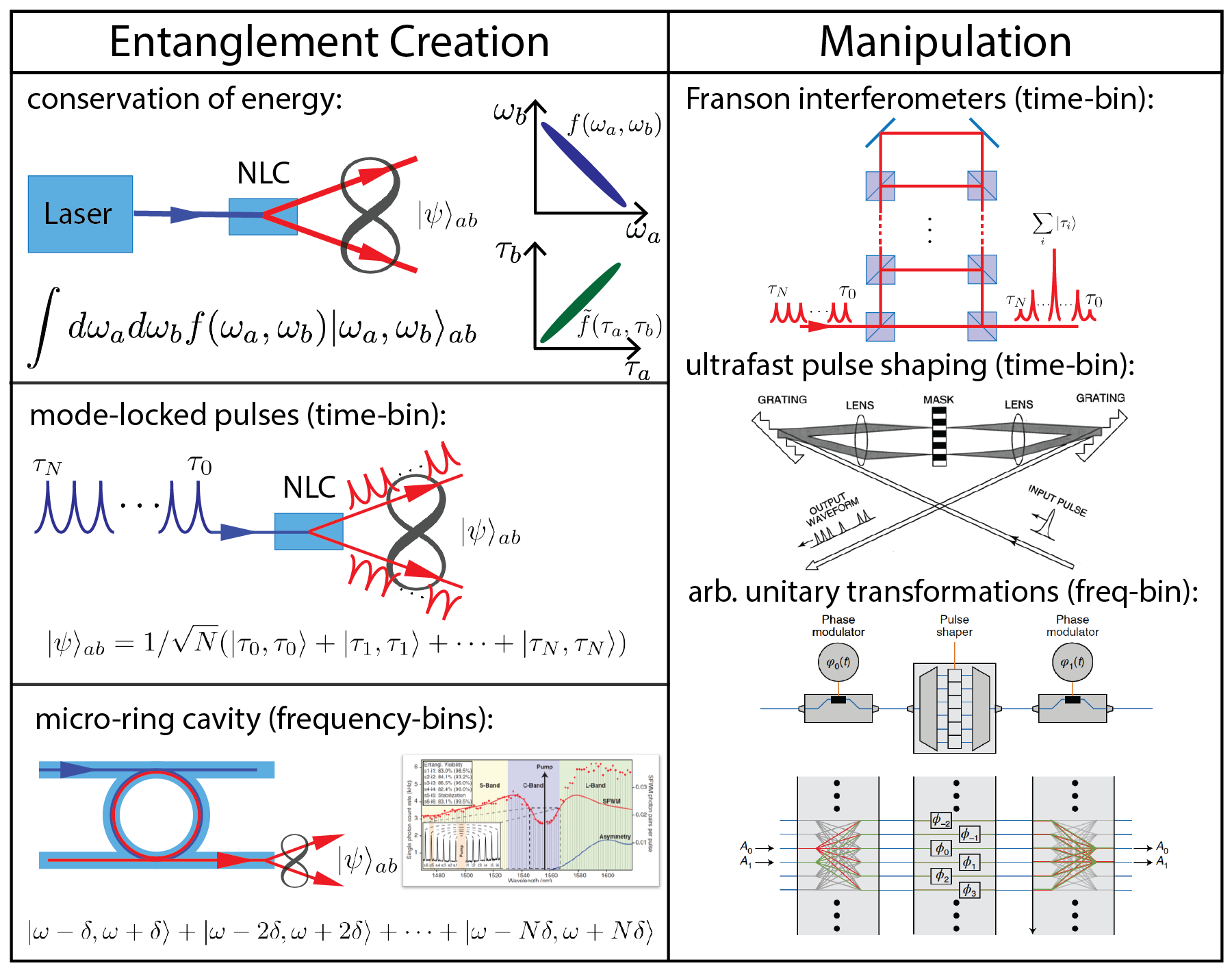}
	\caption{\small{\textbf{Entanglement creation and manipulation concepts for time and frequency bin encoding.} Conservation of energy within the spontaneous-parametric down-conversion process in the non-linear crystal (NLC) can yield very high-dimensionally entangled two-photon quantum states\cite{krenn2014generation} $\ket{\psi}_{ab}$ in principle. The joint-spectral-amplitude (JSA) $f(\omega_a,\omega_b)$ determines the dimensionality. 
	An alternative concept relies on $N$ indistinguishable mode-locked laser pulses. Each laser pulse creates probabilistically one photon pair at time $\tau_i$. Considering N such creation events yields a high-dimensionally entangled two-photon state $\ket{\psi}_{ab}$. A micro-ring cavity can be exploited to create high-dimensionally entangled photon pairs from spontaneous four-wave mixing (SFWM). First, the micro-ring cavity allows for an efficient (SFWM) process due to the high Q-value cavity environment. Secondly, the free-spectral-range (FSR) of this cavity only allows for certain frequency modes to be hosted in this frequency comb, see inset figure taken from\cite{reimer2016generation}. The concept introduced by Franson\cite{franson1989bell} to coherently manipulate time-bin quantum information is depicted above. It relies on several unbalanced Mach-Zehnder type interferometers. In this configuration, it allows producing an equal superposition of all time-bins at the center outgoing bin, as depicted in the graphic. A recently developed new method for arbitrary unitary transformations of frequency bins is shown below, image taken from\cite{kues2019quantum}. An electro-optic-modulator (EOM) populates coherently many neighbouring frequency bins. A consecutive pulse-shaper operating in the complete output space of the EOM introduces arbitrary phase shifts to the separate frequency bins. A final EOM coherently combines all frequency modes and interference among them yields the desired transformation. }}
	\label{fig:time-bin-ent-scheme}
\end{figure*}

One possibility to create entanglement between two photons encoding quantum information in the time-bin domain is to utilize two indistinguishable laser pulses separated by a fixed time difference $\Delta$\cite{brendel1999pulsed}. Each laser pulse can create a photon pair within the non-linear crystal (NLC), see Fig.\ref{fig:time-bin-ent-scheme}. If there is in principle no information available in which of the two pulses the photon pair $a\text{-}b$ has been created, the resulting state can be written as a superposition of the two creation times $(\ket{\tau_1,\tau_1}_\text{ab}+\ket{\tau_2,\tau_2}_\text{ab})/\sqrt{2}$. Note, both pulses emit a photon pair with emission amplitude $\alpha$, such that the quantum state in the Fock basis can be written as $\ket{0,0}_\text{ab}+\alpha\ket{1,1}_\text{ab}+\alpha^2/2\ket{2,2}_\text{ab}+\cdots$ and numbers within the ket vectors refer to the photon occupation number of the respective path mode a or b. It is made sure that the total probability of emitting two pairs simultaneously\footnote{Since these two (or n) events are independent the probability is given by the product, in this case, $\alpha\times \alpha$ or $\alpha^n$.} over the time period $T_\text{coh}$ is well below 0.1 by keeping $\alpha$ small enough. In principle, this scheme can be scaled up to $d$-dimensions by employing a sequence of $d$ indistinguishable and coherent mode-locked pump pulses (see Fig.\ref{fig:time-bin-ent-scheme}) as has been demonstrated in~\cite{DeRiedmatten:2002} for 11 dimensions. 

An alternative method to create HD entangled time-bin photon pairs relies on the intrinsic energy conservation in the SPDC process. By choosing the non-linear optical material, length of the NLC, pump-pulse duration, wavelength, and bandwidth of the pump, signal, and idler beams the joint-spectral-amplitude (JSA see Fig.~\ref{fig:time-bin-ent-scheme}), can be designed to show a high degree of frequency correlations\cite{avenhaus2009experimental,brida2009characterization}. The conjugate variable of frequency is time, and by Fourier transformation of the JSA, we obtain the joint-temporal-amplitude (JTA). Thus by appropriately choosing the parameters mentioned above, a high-dimensional entangled photon pair can be generated. In principle, a very large number of modes within a given time frame is possible. However, as we will see in the following, the restrictions are now imposed by the detection system rather than in the creation part.

In general, the detection of entangled quantum states requires the ability to measure at least in two different bases locally. The main idea to perform such a basis transformation for time-bins is to use an unbalanced interferometer\cite{franson1989bell}. The original proposal was inherently vulnerable to loss (50\%) but can be overcome by using active-switching techniques, as recently demonstrated\cite{vedovato2018postselection}. An unbalanced interferometer creates a superposition between two time-bins $i$ and $i+\Delta$ with an arbitrary phase $\varphi$ as $\ket{\tau_i}+e^{i\varphi}\ket{\tau_{i+\Delta}}$, as described in Fig.\ref{fig:time-bin-ent-scheme}. To get a feeling for how large $\Delta$ is in this unbalanced Franson interferometer, we assume that our mode-locker laser has a repetition rate of 1 GHz. This corresponds to roughly $30$ cm of propagation distance\footnote{assuming the speed of light in vacuum}. Hence, for two-dimensional superpositions of two time-bins created with a 1 GHz repetition laser, a physical propagation difference of 30cm is necessary. If we now imagine a high-dimensional quantum state, e.g. 10-dimensional, it is not only necessary to connect ten such interferometers, but also the distances increases to several meters. To overcome this limitation, a theoretical dimensionality bound allows to measure only next and over-next neighbouring time-bins\footnote{meaning superpositions of $j\in\{1,2\}$ $\ket{i}+\ket{i+j}$} to put a lower bound to the entanglement dimensionality being created\cite{martin2017quantifying}. In this work the authors observed an 18-dimensionally entangled quantum state. 

Another possibility to perform transformations of the quantum information stored in the time-bin domain is inspired by techniques developed in the ultrafast optical pulse shaping community\cite{weiner1988high,yelin1997adaptive,weiner2000femtosecond}. The basic idea is to utilize a 4f-nondispersive pulse shaper with a programmable spatial-light-modulator (SLM) inserted, see Fig.\ref{fig:time-bin-ent-scheme}. This technique is used to shape the temporal distribution of entangled photon pairs\cite{pe2005temporal}.  

To overcome timing constraints from detectors plus counting electronics and thus enabling ultra-fast timing detection in the picosecond regime, non-linear optical approaches based on coherent sum-frequency generation (SFG) can be employed\cite{pe2005temporal,donohue2013coherent}. In a recently published experiment, this technique was perfected and allowed the direct characterization of the spectral and temporal properties of energy-time entangled photon pairs on the sub-picosecond timescale\cite{maclean2018direct}.

%An alternative method to measure two bases using time-bins includes the concept of a \textit{time lenses}. A time lens is based on a non-linear effect and effectively mimics the behaviour of a spatial lens.

Recent developments in upcoming on-chip schemes enable the creation of high-dimensionally entangled photon pairs in the form of discrete frequency bins\cite{ramelow2009discrete,olislager2010frequency,bernhard2013shaping}. This new approach allows for versatile and high-quality sources of two and multi-photon quantum states in higher-dimensions\cite{kues2017chip,reimer2016generation,imany201850}. The most commonly used techniques are based on integrated Kerr frequency combs\cite{kues2019quantum}. Due to energy conservation in the spontaneous four-wave mixing (SFWM) process inside the micro-ring cavity, perfectly anti-correlated frequency-bin entangled qudits are created. The number of dimensions achievable is determined by two factors. First, the phase-matching range of the SFWM process sets the overall bound for the correlated frequency range. For currently used materials\footnote{E.g. Hydex, a special CMOS compatible material with a low $\chi^3$ non-linear coefficient but capable of hosting non-linear optical effects such as SFWM in a high Q-value cavity environment\cite{moss2013new}. Alternatively, also silicon nitride (SiN) based micro-ring cavities\cite{imany201850} are used.}, a bandwidth of roughly $100$nm at 1550nm telecom wavelength can be achieved. The discrete frequency bins are now directly created within the micro-ring cavity. Their spacing is given by the free-spectral-range (FSR) of the micro-ring resonator. Typical values of the FSR are 200GHz with a line-width of roughly 800MHz\cite{reimer2016generation}. These values allow for entanglement up to several tens of frequency bins, see Fig.\ref{fig:time-bin-ent-scheme}. Most importantly, the large FSR also enables the use of off-the-shelf telecom equipment for accessing and manipulating (individually or collectively) the quantum information encoded in frequency bins.

This promising source technology (in combination with recent advances in creating quantum gates for discrete frequency bins such as generalized splitters in higher-dimensions\cite{lu2018electro}) makes frequency encoded qudits a promising platform. As displayed in Fig.\ref{fig:time-bin-ent-scheme}, a pulse-shaper sandwiched between two electro-optical modulators (EOM) is used to perform a frequency tritter. The first EOM scatters any input mode coherently over its neighbouring modes. This, however, introduces unwanted loss as it scatters some of the modes outside of the desired state space. To overcome this limitation, a pulse shaper operating in all possible output modes of the EOM can apply arbitrary phases to each frequency bin. A final EOM then coherently recombines all modes. This procedure allows for any arbitrary unitary transformation. An important example is shown in Fig.\ref{fig:time-bin-ent-scheme}. There, a frequency-tritter\footnote{A tritter is a generalized splitter for three dimensions.} or Quantum-Fourier transform (QFT) is demonstrated. It takes any of the three frequency bins as input and creates an equal and coherent output distribution of all three frequency bins\footnote{including phases of $exp(\frac{n}{3}\dot 2\pi i)$ for unitarity}. The high-fidelity operation of up to 99.9\% with standard telecom equipment is very promising. A remaining problem is the relatively large coupling loss of 3-4dB per element. This loss could be overcome in the future using on-chip techniques\cite{lu2018electro}.

An equivalent device to the even/odd sorter for OAM also exists for frequency bins\cite{olislager2014creating}. It is called \textit{optical frequency interleaver} with two input and two output channels. Thus it allows to sort frequency bins with a spacing $\Delta$ into even $\ket{f_0+2m\Delta}$ and odd $\ket{f_0+(2m+1)\Delta}$ frequency modes with respect to some central frequency $f_0$ and $m\in\mathbb{N}^+$. Since the frequency interleaver is a multi-input/multi-output device, this readily allows to create special types of quantum gates using concepts originally found for OAM ranging from high-dimensional quantum gates~\cite{babazadeh2017high,gao2019arbitrary} to genuine high-dimensional and multi-photon entanglement\cite{malik2016multi,erhard2018experimental} and even to high-dimensional controlled quantum gates\cite{gao2019computer}.

\begin{figure*}[t!]
\centering
\includegraphics[width=.85\linewidth]{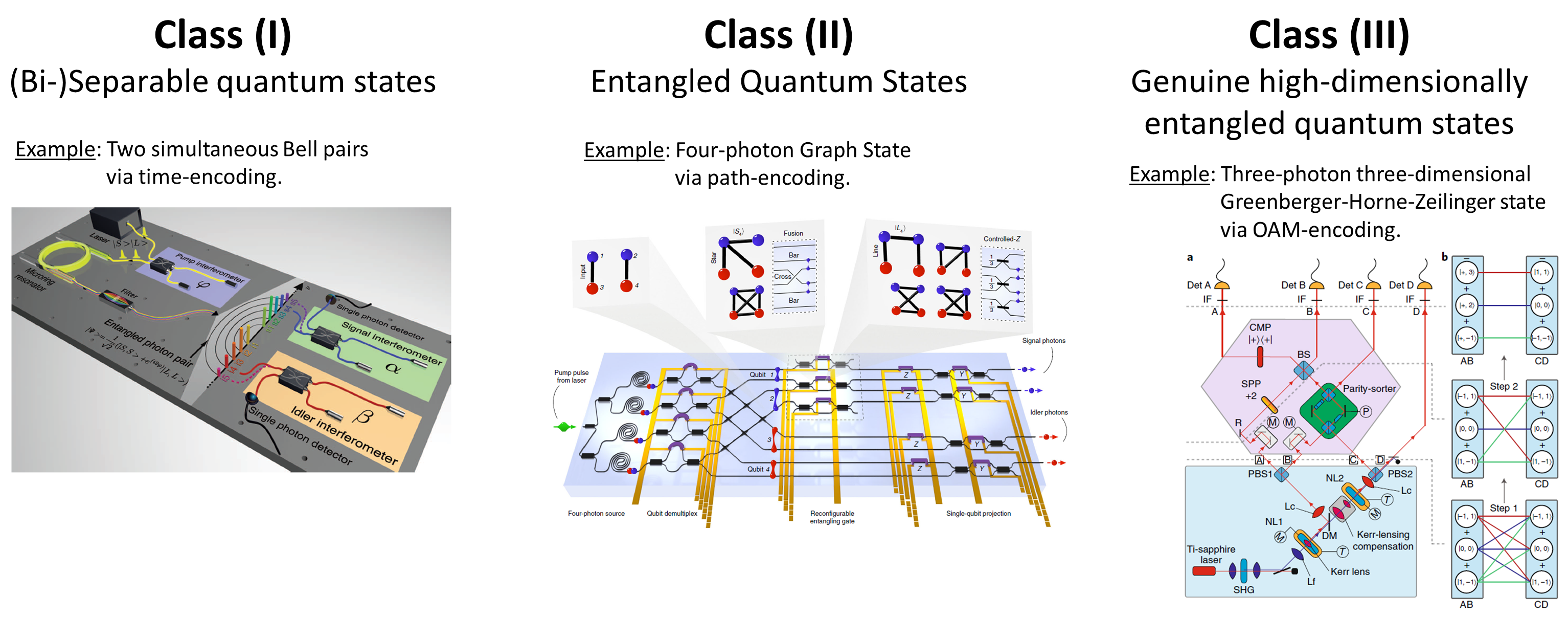}
\caption{\textbf{Examples of the three classes of multi-particle entanglement in terms of dimension.} Class (I) contains multi-photonic, separable quantum states, such as ref. \cite{reimer2016generation}, which created two pairs of time-entangled photon pairs. Class (II) contains genuinely multi-photonic entangled states, such as ref.\cite{adcock2019programmable}, which produced a 4-photonic graph state in the path degree-of-freedom on a chip. Finally, Class (III) contains genuinely high-dimensional multi-photonic states, such as ref.\cite{erhard2018experimental} which produced a 3-photonic 3-dimensional GHZ state. }
\label{fig:multiphoton-classes}
\end{figure*}

The time-bin degree of freedom is ideally suited for HD-QKD. One possibility to implement an HD-QKD protocol is to use visibility measurements in Franson type interferometers\cite{ali2007large}. Another experiment recently achieved 2.7 Mbit/second secure key rate over a transmission distance of 20km in an optical fiber\cite{zhong2015photon}. In this experiment, the security of the HD-QKD protocol is ensured by measurements of the Franson visibility. Their security analysis includes all Gaussian attacks (such as beam-splitter attacks). However, intercept-and-resend attacks are not considered. Unconditional secure HD-QKD schemes have been proposed using Franson in combination with conjugate Franson interferometry\cite{zhang2014unconditional}, for example.
Dispersion cancellation systems are commercially available. Hence the long-distance distribution of discrete frequency-bins via standard optical fibers up to 40km is possible and has been demonstrated.

A recent study investigated the noise resistance of higher-dimensional entangled photon pairs. Using time-bins, the authors experimentally certify entanglement with a noise-fraction of 92\% for time-bin entanglement in 80-dimensions \cite{ecker2019entanglement}. In the same study, a parallel experiment using OAM encoding achieved a noise robustness of 63\% in a 7-dimensional space.

\subsection{Combining and converting different DoFs}
A powerful technique to achieve high-dimensional quantum states is to utilize several DoFs of single photons simultaneously. One possibility is to use the polarisation (p), OAM (o) and time-frequency (t-f) DoFs being entangled in two and three dimensions yielding\cite{barreiro2005generation}
\begin{eqnarray}
&\ket{\Phi}_\text{p}\otimes\ket{\Phi}_\text{o}\otimes\ket{\Phi}_\text{t-f}=\underbrace{(|H,H\rangle+|V,V\rangle)}_\text{polarisation}\otimes\\\nonumber
&\underbrace{(|0,0\rangle+|1,-1\rangle+|-1,1\rangle)}_\text{OAM}\otimes\underbrace{(|s,s\rangle+|l,l\rangle)}_\text{time},
\end{eqnarray}
with $H,V$ denoting horizontal and vertical polarisation, $s,l$ describing short and long of creation-times of the photon pair and numbers within the ket-vectors refer to quanta of OAM in units of $\hbar$. This state can be rewritten in terms of these 12 orthogonal states $\ket{H,0,s},\ket{H,0,l},\cdots,\ket{V,1,l}$ and results in a $(2\times3\times2=12)$-dimensionally entangled quantum state in three DoFs simultaneously. Such states can be created by employing techniques described above. The advantage is that for hybrid DoFs there exist deterministic \cnot-gates, for example, which is important for applications such as superdense coding\cite{barreiro2008beating} or superdense quantum teleportation\cite{graham2015superdense}.

A very interesting idea is to convert one DoF to another. In such an approach, entanglement created in path DoF, for example, can directly be converted to OAM entanglement\cite{fickler2014interface}. This is achieved by using the mode sorter between OAM and path\cite{berkhout2010efficient,mirhosseini2013efficient} in reverse. To generate a three path entangled two-photon state, an NLC produces pairs of photons that illuminate a three slit mask. The consecutive mode sorter transforms the path information to OAM states while maintaining the entanglement encoded in the path DoF. Such interfaces allow to create and manipulate quantum states on-chip using the path DoF and then connect to distant chips via a quantum link encoded in OAM DoF.

\section{Experimental multi-photon entanglement in high-dimensions}

\subsection{Multi-Photon Entanglement in high Dimensions}
Entanglement with multiple (n>2) photons in higher (d>2) dimensions can have complex structures even in the case of pure states \cite{huber2013structure,huber2013entropy}.  
The Schmidt-Rank-Vector (SRV) allows to classify these structures and is defined as the collection of Schmidt-Ranks of all bi-partitions such that
\begin{equation}
    \text{SRV}=\{r_1,r_2,\cdots,r_k\},
\end{equation}
holds, with $r_i$ denoting the rank of the reduced density matrix ${r_i=\text{Rk}[\text{Tr}_{\bar{i}}(|\psi\rangle\langle\psi|)]}$ and $k$ represents the number of possible bi-partitions of a $N$-partite quantum system given by ${k=2^{N-1}-1}$.
Most of the experiments involving multiple photons in a high-dimensional degree-of-freedom can be distinguished into the following three cases:

\begin{enumerate}[(I)]
	\item \textit{Entanglement involving many photons and in a high-dimensional degree of freedom}: Such states involve more than two photons, but can be bi-separable. An example is $\ket{\psi}=\ket{\phi_{A,B}}\otimes\ket{\phi_{C,D}}$. Their Schmidt-Rank-Vector has some entries being 1.
	\item \textit{Genuine multi-photonic entanglement in a high-dimensional degree of freedom}: These states are not separable, but their entanglement is not high-dimensional (even though encoded in a high-dimensional space). The state is not separable, and some or all parts are two-dimensionally entangled. An example would be a 4-photon 2-dimensional GHZ state in the path degree of freedom.
	\item \textit{Genuine multi-photonic high-dimensional entanglement}: All photons are entangled in more than two dimensions.
\end{enumerate}
See the examples in Fig.\ref{fig:multiphoton-classes} and Details in Table~\ref{tab:multiphotonhighdim}.

\subsubsection{Path}
Many multi-photon experiments in the path degree-of-freedom \cite{spring2012boson,broome2013photonic,tillmann2013experimental,crespi2013integrated,bentivegna2015experimental,wang2017high,zhong201812,paesani2019generation,wang2019boson}, have recently been motivated by Aaronson\&Arkhipov's BosonSampling proposal \cite{aaronson2011computational}, which shows that linear optics can perform transformations that are difficult for classical computers. The purpose of these experiments has not been to produce well-defined entangled states, and the measurements did not measure the full state but investigated the probability distribution between different modes. 

In 2019, the first genuine multiphoton entangled state has been generated and measured on a programmable chip \cite{adcock2019programmable}. The authors demonstrated different types of graph states, among them the 4-photon Star graph $\ket{S_4}$, which is locally equivalent to a 2-dimensional GHZ state. Thereby the authors have demonstrated the first \textit{class (II)} entangled state in the path degree of freedom. The photon pairs (with signal and idler wavelength at 1539nm and 1549nm, respectively) have been created directly on the chip, which significantly improves the stability, necessary for such complex experiments. The key component for producing genuine multiphoton entanglement is a reconfigurable, postselected entangling gate which exploits Hong-Ou-Mandel interference\cite{hong1987measurement} to remove the \textit{which-crystal information} between the pairs from different origins. The chip contains four arbitrarily tuneable single-qubit projections, which allows for the measurement of arbitrary multi-photon correlations in coincidences. The fidelity was 78\%, and count rates were in the order of 5.7 mHz (20 per hour). The count rates are similar to those in the first photonic multiphoton experiments, and the authors explain technical improvements which could lead to pushing four-photon rates from the mHz to the kHz regime. A main future objective is the reduction of photon loss, currently in the range of 19.3 dB for the presented device. Using published values of record component efficiencies, the authors estimate a potential four-photon count-rate improvement of a factor 5 million.

\begin{table*}[t!]
  \centering
  \begin{tabular}{c c c c c c}
  \toprule
     & \textbf{State} & \textbf{Class} & \textbf{Fidelity} & \textbf{Counts/h} & \textbf{Year}\\
    \midrule
    \midrule
    \textbf{Path} & 
        \begin{tabular}{@{}c@{}}BosonSampling\cite{spring2012boson,broome2013photonic,tillmann2013experimental,crespi2013integrated,bentivegna2015experimental,zhong201812,paesani2019generation} \\ 4n Star Graph\cite{adcock2019programmable}\end{tabular}
     & 
        \begin{tabular}{@{}c@{}}\textit{class (I)} \\ \textit{class (II)}\end{tabular}
     & 
        \begin{tabular}{@{}c@{}} - \\ 78\%\end{tabular}
     & 
        \begin{tabular}{@{}c@{}} - \\ 20\end{tabular}
     & 
        \begin{tabular}{@{}c@{}} 2012-2019 \\ 2019\end{tabular}        \\
    \midrule
    \textbf{Spatial Modes} & 
        \begin{tabular}{@{}c@{}}4n Dicke State\cite{hiesmayr2016observation} \\ SRV=(3,3,2)\cite{malik2016multi} \\ 3d GHZ\cite{erhard2018experimental} \end{tabular}
    & 
        \begin{tabular}{@{}c@{}}\textit{class (II)} \\ \textit{class (II)} \\ \textit{class (III)} \end{tabular}
    & 
        \begin{tabular}{@{}c@{}}62\% \\ 80,1\% \\ 75,2\% \end{tabular}
    & 
        \begin{tabular}{@{}c@{}}720 \\ 55 \\ 4 \end{tabular}
     & 
        \begin{tabular}{@{}c@{}} 2016 \\ 2016 \\ 2018\end{tabular}        \\ 
    \midrule
    \textbf{Time/Frequency} & 
        \begin{tabular}{@{}c@{}}2 Bell Pairs\cite{reimer2016generation} \\ BosonSampling\cite{he2017time}\\3n W state\cite{fang2019three} \end{tabular}
     & 
        \begin{tabular}{@{}c@{}}\textit{class (I)}\\\textit{class (I)} \\ \textit{class (II)}\end{tabular}
     & 
        \begin{tabular}{@{}c@{}} 64\%\\ - \\ 90\%\end{tabular}
     & 
        \begin{tabular}{@{}c@{}} 600 \\ -\\75\end{tabular}
     & 
        \begin{tabular}{@{}c@{}} 2016 \\ 2017\\ 2019\end{tabular} \\
    \midrule
    \textbf{Hybrid Entanglement} & 
        18 qubit GHZ\cite{wang201818}
     &  \textit{class (II)}
     & 70.8\% 
     & 200 & 2018\\
  \bottomrule
  \end{tabular}
  \caption{Summary of multi-photon experiments in high-dimensional degrees of freedom.}
  \label{tab:multiphotonhighdim}
\end{table*}

\subsubsection{Spatial modes of photons}
The first multi-photon entangled state with spatial modes of light was created in 2016\cite{hiesmayr2016observation}, which was a state of \textit{class (II)}. The authors used double-pair emissions of an SPDC crystal, which has then been probabilistically split using three beam splitters -- in a similar way as the state in eq.(\ref{frequencybin3W}). The resulting state is a Dicke state in the form 
\begin{align}
	\ket{\Psi}&=\frac{1}{\sqrt{6}}(\ket{0,0,1,1}+\ket{0,1,0,1}+\ket{1,0,0,1}+\nonumber\\
	&+\ket{0,1,1,0}+\ket{1,0,1,0}+\ket{1,1,0,0} \large),
  \label{oam4photon}
\end{align}
where $\ket{0}$ and $\ket{1}$ stand for two different OAM modes. This state falls into \textit{class (II)}. The authors demonstrate a fidelity of 62\% with a count rate of roughly 0.2 Hz. 

Later in the same year, a different group of authors demonstrated a three-photon entangled state, where part of the state was entangled in three dimensions\cite{malik2016multi}, $\ket{\psi}=\frac{1}{\sqrt{3}}\left(\ket{0,0,0}+\ket{1,1,1}+\ket{2,2,1}\right)$. While still not fully high-dimensionally entangled (thus still a \textit{class (II)} entangled state), it is the first time that part of a multi-photon entanglement is higher dimensional. The fidelity of the state was 80.1\% with a count rate of approximately 15mHz.

In 2018, the first genuine high-dimensional multi-particle entanglement has been created\cite{erhard2018experimental}, in form of a three-dimensional GHZ state
\begin{align}
	\ket{\Psi}=\frac{1}{\sqrt{3}}\left(\ket{0,0,0}+\ket{1,1,1}+\ket{2,2,2}\right).
  \label{OAM3dimGHZ}
\end{align}
The experimental setup, which has been discovered by \melvin \cite{krenn2016automated}, consists of two sources of SPDC crystals which produce three-dimensional photon pairs. The exploitation of a multi-port (MP) that coherently manipulates several photons simultaneously in higher dimensions allows to removes all cross-correlation terms. This experiment demonstrated for the first time \textit{class (III)} entanglement. The state fidelity was 75.2\% with a count rate of 1.2 mHz (roughly 4 counts per hour).

It is important to note that all concepts are translatable between the different degrees of freedom. It means, for example, that the multiport (MP) could directly be employed for path- or time-bin encoding of high-dimensional quantum information.

\subsubsection{Discretized Time and Frequency modes}
In addition to efforts into BosonSampling with time-encoding \cite{he2017time}, only two experiments have demonstrated entanglement with more than two photons with discretized time-bins \cite{reimer2016generation}. In the first one, a quantum frequency comb is generated, which contains a large number of discrete, equally spaced frequency lines (roughly 100 bins within 100 nm) distributed symmetrically around the pump wavelength of 1550nm. Each symmetric pair of frequency lines can be occupied by photon pairs. The authors now use these pairs of bins for creating time-bin entanglement in the form $\ket{\phi_n}=\frac{1}{\sqrt{2}}\left(\ket{S_i,S_s}+e^{i\phi_n}\ket{L_i,L_s}\right)_n$, for the $n$-th frequency pair symmetrically arranged around the pump wavelength.

Next, they show that their quantum frequency comb allows for the occupation of two pairs at the same time in a coherent way. Specifically, they generate the state
\begin{align}
	\ket{\Psi}&=\ket{\psi_1}\otimes\ket{\psi_2}\nonumber \\
	&=\frac{1}{2}(\ket{S_{s_1},S_{i_1},S_{s_2},S_{i_2}}+e^{i\phi}\ket{S_{s_1},S_{i_1},L_{s_2},L_{i_2}} +\nonumber \\
	&+e^{i\phi}\ket{L_{s_1},L_{i_1},S_{s_2},S_{i_2}} + e^{i2\phi}\ket{L_{s_1},L_{i_1},L_{s_2},L_{i_2}}  )
  \label{timebin4photon}
\end{align}
By changing the phase $\phi$, the authors show that they created indeed a coherent 4-photon state. This state falls into the \textit{class (I)}, because it is separable. It reached fidelities of 64\% with a count rate of roughly 0.17 Hz (600 per hour). The authors argue that loss reductions could lead to kHz four-photon rates. To create genuine multiphotonic entanglement, one needs to delete the \textit{which-frequency-bin information}.

Very recently, a three-photon W state has been created and verified using discretized frequency\cite{fang2019three}, which is \textit{class (II)} entangled. The state was generated by two pairs from spontaneous four-wave mixing in a polarisation-maintaining fiber, where each pair consists of a blue (694nm) and a red (975nm) photon. The four photons are then probabilistically split to four detectors. One trigger photon then indicates the existence of a W state in the other three detectors (in post-selection). The specific state is
\begin{align}
	\ket{\Psi}=\alpha\ket{\psi_0}+\gamma\left(\ket{B,W_1}+\ket{R,W_2}\right)
  \label{frequencybin3W}
\end{align}
with $\ket{W_{1/2}}$ being W state
\begin{align}
	\ket{W_1}&=\frac{1}{\sqrt{3}}\left(\ket{R,R,B}+\ket{R,B,R}+\ket{B,R,R} \right)\\
	\ket{W_2}&=\frac{1}{\sqrt{3}}\left(\ket{B,B,R}+\ket{B,R,B}+\ket{R,B,B} \right),
  \label{frequencybin3W2}
\end{align}
where $\ket{R}$ ($\ket{B}$) stands for a red (blue) photon. Using imbalanced interferometers, the authors are able to verify the coherence between the individual terms, and estimate a fidelity of roughly 90\%, with a count rate of 20 mHz (roughly 75 counts per hour). 

\subsubsection{Multiple degrees-of-freedom}
The technologies can not only be translated between different degrees of freedom, but quantum entanglement can be exploited within different degrees in many photons as well. This allows, for example, to encode more than one qubit per photon. In 2018, an 18 qubit entangled state has been created, encoded in six photons with three degrees of freedom (polarisation, path and spatial modes) \cite{wang201818}. 

For the generation, the authors start with high-fidelity six-photon states in the form $\ket{\psi}=\frac{1}{\sqrt{2}}\left(\ket{H,H,H,H,H,H}+\ket{V,V,V,V,V,V} \right)$, where $H$ and $V$ stand for horizontally and vertically polarized photons. Afterwards, each of the six photons undergoes a transformation $\ket{H} \to \ket{H}\ket{U}\ket{R}$ and $\ket{V} \to \ket{V}\ket{D}\ket{L}$, where $U$ and $D$ stand for path labels (\textit{up} and \textit{down}), while $R$ and $L$ stand for right- and left-circular parity of the photon's OAM. This leads to an 18 qubit entangled GHZ state. The fidelity is measured to be 70.8\% and the six-fold count rate is 55 mHz (roughly 200 counts per hour).

With the techniques described in the previous chapter, it is conceivable that this technique could be extended to time-bins, which would allow for 24 qubits. Using the recently demonstrated 12-photon entanglement source\cite{zhong201812}, 48-qubit entangled GHZ states seem feasible with current technology.

\subsection{Quantum Teleportation in high Dimensions}

\begin{figure}[ht]
\centering
\includegraphics[width=0.9\linewidth]{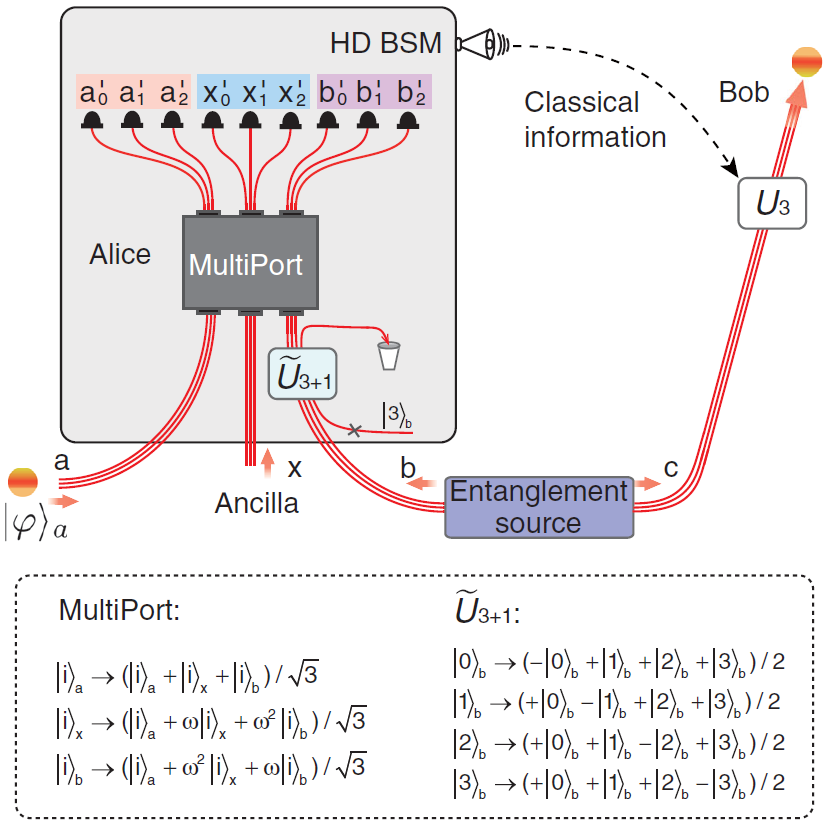}
\caption{Concept for performing 3-dimensional teleportation. One photon from the entangled pair is transformed in an extended, 4-dimensional space, $\tilde{U_{3+1}}$. Afterwards, a multiport is used to mix the entangled photon, the teleportee and an ancillary photon to perform a 3-dimensional Bell state measurement. Importantly, the concept is entirely general and can be implemented in any degree-of-freedom to perform high-dimensional teleportation. This concept has been implemented in the path degree-of-freedom, to teleport the first three-dimensional quantum state\cite{luo2019quantum}.}
\label{figureSRVlist}
\end{figure}

Quantum teleportation, the \textit{disembodied} transmission of unknown quantum states, is one of the most fascinating processes allowed by quantum mechanics. Discovered in 1993 as a conceptual curiosity\cite{bennett1993teleporting}, it has become a cornerstone in various quantum applications, such as quantum computation and long-distance quantum key distribution networks. To teleport a system from A to B, one has to share an entangled photon pair between these two locations. A Bell-state measurement then projects the two particles at A into a joint state -- thus removing their identities. The classical information of the joint measurement outcome is then sent to B, where an outcome-dependent local transformation recreates the initial quantum state. 

The first experimental demonstration of quantum teleportation has transmitted the polarisation information of a single photon, which resembles a two-state quantum system\cite{bouwmeester1997experimental}. Stretching the idea to larger systems, researchers found ways to teleport the quantum information of multiple particles at the same time \cite{zhang2006experimental} and multiple properties of a single particle\cite{wang2015quantum}.

The final obstacle, quantum teleportation of high-dimensional systems has turned out to be conceptually more difficult: The major challenge in teleporting high-dimensional photonic quantum states has been identified by Calsamiglia already in 2002\cite{calsamiglia2002generalized}. He showed that high-dimensional Bell-state measurements, with linear optical components, requires additional ancillary particles. The key insight is that with linear optics only, it is impossible to distinguish a single Bell state unambiguously from the other $d^2$ -- except for $d=2$. Several theoretical approaches have been developed to overcome Calsamiglia's no-go theorem\cite{goyal2013teleporting, goyal2014qudit, zhang2019arbitrary, zhang2019quantum}. Additional challenges come from the fact that the teleportation fidelity needs to be beyond $F=2/3$ in order to demonstrate genuine three-dimensional teleportation. Fidelities below 50\% can be achieved with classical techniques, and fidelities between 50\% and 66.6\% can be achieved using qubit systems.

In 2019, two experiments have been reported which demonstrate teleportation using three-dimensional path-encoding of a photon\cite{luo2019quantum, hu2019experimental}. In ref.\cite{luo2019quantum}, a three-dimensional Bell state measurement has been experimentally demonstrated that exploits quantum-Fourier transformation and can be generalized to arbitrary dimensions. Their approach is optimal in terms of required additional photons. The experiment required four photons (the \textit{teleportee}, a three-dimensionally entangled photon pair, and an ancillary photon to overcome Calsamiglia's no-go theorem). The authors demonstrate count rates of 110 mHz (roughly 400 counts per hour), and a teleportation fidelity of 75\%, well beyond both the classical and qubit bound.

The experiment in ref. \cite{hu2019experimental} demonstrated teleportation using a Bell state measurement which requires an entangled ancillary pair of photons -- making the demonstration a six photon experiment. The authors report a count rate of 2 mHz (roughly ten counts per hour) and a teleportation fidelity of 63.8\%. This value is well beyond the classical bound for teleportation.

Both experiments demonstrate high-quality long-time stability of their (not integrated) experiments. The authors of ref.\cite{hu2019experimental} demonstrate that their interferometers retain remarkable interference visibilities beyond 98\% for 48 hours. The conceptual ideas and technologies can be transferred to and combined with technology in other high-dimensional degrees of freedom, which would enable the employment of these techniques over large distances (as it has been achieved for qubit systems\cite{ren2017ground}), and follow the dream of teleporting the entire quantum information of a quantum system.

\subsubsection{Applications in Entanglement Swapping}
Extending the teleportation scheme to a situation where the teleportee photon itself is entangled with another photon leads to entanglement swapping\cite{zukowski1993event}. The intriguing fact in this scenario is that two photons become entangled that never interacted before nor ever shared a common past. Entanglement swapping has become an important fundamental concept, for example, to overcome long distances in quantum networks \cite{briegel1998quantum, yuan2008experimental} or in fundamental experiments regarding entanglement\cite{pan1998experimental,hensen2015loophole}. Generalizing this concept to more complex entanglement structures in higher-dimensions has not fully been achieved yet, but some important intermediate results have been reported.

%Teleporting one photon of an entangled pair leads to entanglement between photons that never shared a common path, which is called entanglement swapping\cite{zukowski1993event}. It has become an important fundamental concept, for example, to overcome long distances in quantum networks \cite{briegel1998quantum, yuan2008experimental} or in fundamental entanglement experiments \cite{pan1998experimental,hensen2015loophole}. Generalizing this concept to more complex entanglement in higher-dimensions has not fully been achieved yet, but some important intermediate results have been reported.

In 2017, the first entanglement swapping experiment with a high-dimensional degree of freedom was reported\cite{zhang2017simultaneous}. There the authors perform a Bell-state projection onto a large number of 2-dimensional subspaces. Conditioned on two-fold detections, they know that two initially independent photons are now in an incoherent superposition of many two-dimensionally entangled states, in the form 
\begin{align}
	\rho&=c_1 \ket{\psi^-_{-1,1}}\bra{\psi^-_{-1,1}} + c_2 +\ket{\psi^-_{-2,2}}\bra{\psi^-_{-2,2}}  \nonumber\\ 
	&+c_3 \ket{\psi^-_{-2,1}}\bra{\psi^-_{-2,1}} + c_4 \ket{\psi^-_{-1,2}}\bra{\psi^-_{-1,2}} \nonumber\\ &+c_5\ket{\psi^-_{1,2}}\bra{\psi^-_{1,2}} + c_6 \ket{\psi^-_{-1,-2}}\bra{\psi^-_{-1,-2}}.
  \label{OAM2dimswapping}
\end{align}

The authors report average fidelities of the two-dimensionally entangled swapped subspaces of $F=80\%$ (background subtracted; raw: $F=57\%$) for the six 2-dimensionally entangled systems. They also explain in detail that their system is not high-dimensionally entangled, as one would need a three-dimensional Bell state measurement.

An interesting application of their method lies in Ghost imaging\cite{bornman2019ghost}. In conventional Ghost imaging \cite{moreau2019imaging}, the object never interacts with the photon used to image it -- but only with its correlated partner photon. In the demonstrated scheme, the photon interacting with the object does not even share a common past with the photon used for imaging. Thus it is conceptually interesting, whether quantum or classical correlations are required for this task.

\section{Future Directions for photonic high-dimensional Entanglement}
The proof-of-principle experiments detailed in this review define the current technological and conceptual state-of-the-art. Many essential challenges are awaiting in the next few years, to enhance conceptual understanding and practical applicability of high-dimensional multi-photonic entanglement.

On the technological side, highly-efficient or deterministic single-photon or photon-pair sources are necessary to increase count rates of multi-photon experiments \cite{wang2019towards}. The development of near-unity photon detectors with small timing jitter will be significant for scaling up time-bin entangled states. The employment of efficient multi-outcome detection schemes (e.g. with detector arrays) will be necessary to harness the high-dimensional information stored in single photons. Scaling integrated optics to more modes, and in particular, multi-photon generation on-chip will be essential for path-encoding schemes\cite{wang2019integrated}. The access to complex entangled states gives rise to interesting properties, such as absolutely maximally entangled quantum states \cite{goyeneche2015absolutely,cervera2019quantum}, and their peculiar features can not be investigated in laboratories. A critical question is how to control and manipulate entangled multi-photonic states experimentally. It remains open whether some of the unconventional methods that have been demonstrated successfully for single quDits \cite{popoff2010measuring,fickler2017custom, brandt2019high} will still be feasible when many particles are concerned.

% Wir könnten hier noch erwähnen, dass event ready Bell tests in höheren dimensionen es erlauben könnten lokal realistische Verletzung auf astronomischen distanzen zu ermöglichen (zB die satelliten mission der chinesen an der lagrange punkten). Dies könnte einflüsse der Gravitation auf Verschränkung ermöglichen. Auch experimentelle Umsetzungen von Process-Matrices und incausal structures wären interessant. Vor allem deren Experimentelle umsetzung, welche vielleicht mit hoch-dimensionalen systemen gelingen könnte.

In basic research on high-dimensional many-body entanglement, several questions remain unanswered. For example, the generalisation of high-dimensional multi-photonic all-versus-nothing violations of local realism is a way to extend our understanding of the severe difference between our classical worldview and predictions of quantum mechanics. Here, of particular importance are generalisations of the GHZ argument \cite{ryu2013greenberger,PhysRevA.89.024103,lawrence2014rotational, lawrence2019many}, and its application to asymmetrically entangled states that only appear when both the number of particles and dimension is beyond two \cite{huber2013structure}. This involves the understanding of whether non-hermitian measurements are necessary to find the most extensive violations, as they have appeared in recent results. It is an open question of what the most efficient protocols for high-dimensional quantum teleportation are, and how multiple (high-dimensional) degrees of freedom can be experimentally teleported. This will be essential for the philosophically appealing goal of teleporting the entire quantum information of a single photon. 

Of course, usually the most exciting advances are \textit{unknown unknowns}, thus cannot be predicted. In the light of the enormous progress in the experimental capabilities of high-dimensional multipartite quantum entanglement in just the last five years, we would like to encourage and invite theoretical and experimental quantum scientists to explore and exploit the hidden potential of complex, high-dimensional many-body quantum entanglement.

\section*{Acknowledgements}
We thank Sebastian Ecker and Lukas Bulla for insightful discussions related to time-bin entanglement. This work was supported by the Austrian Academy of Sciences ({\"O}AW), University of Vienna via the project QUESS and the Austrian Science Fund (FWF) with SFB F40 (FOQUS). ME acknowledges support from FWF project W 1210-N25 (CoQuS). MK acknowledges support from FWF via the Erwin Schr\"odinger fellowship No. J4309. 

%\newpage

\bibliography{sample}
\end{document}

%% file: preamble.tex
\usepackage{amsthm}
\usepackage{mathtools}
\usepackage{physics}
\usepackage{xcolor}
\usepackage{graphicx}
\usepackage[left=23mm,right=13mm,top=35mm,columnsep=15pt]{geometry} 
\usepackage{adjustbox}
\usepackage{placeins}
\usepackage[T1]{fontenc}
\usepackage{lipsum}
\usepackage{csquotes}